\documentclass[runningheads]{llncs}
\usepackage[T1]{fontenc}
\usepackage{graphicx}

\usepackage{float}
\usepackage{tabularx}
\usepackage{multirow}
\usepackage{listings}
\usepackage{xcolor}
\usepackage{soul}
\usepackage{amsmath}
\usepackage{amssymb}
\usepackage{graphicx}
\usepackage{lscape}
\usepackage{subcaption}
\usepackage{url}
\usepackage{hyperref}
\usepackage{enumitem}
\usepackage{comment}

\usepackage{orcidlink}
\usepackage{bbding}

\begin{document}
\title{From Security Events to Conflict States: A Three-layer Cyber Defense Scenario Model for Enhanced Cyber Situational Awareness}


\titlerunning{A Three-layer Cyber Defense Scenario Model}

 \author{
 Miguel Requena Micó\inst{1}\orcidlink{0009-0001-3408-2791}
 \and
 Mario Fernandez-Tarraga\inst{1}\orcidlink{0009-0009-6403-3243}
 \and
 Daniel Díaz-López\inst{1}\Envelope\orcidlink{0000-0001-7244-2631}
 \and
 Sergio López Bernal \inst{1}\orcidlink{0000-0003-1869-1965}
 \and
 Gregorio Martínez Pérez\inst{1}\orcidlink{0000-0001-5532-6604}
 }
 \authorrunning{M. Requena et al.}
 %
 \institute{
 Departamento de Ingeniería de la Información y las Comunicaciones, Universidad de Murcia, Murcia 30100, Spain\\
 \email{\{miguel.requenam, mario.fernandezt, danielorlando.diaz,  slopez, gregorio\}@um.es}
 }
\maketitle              
%

%

\begin{abstract}
Cyber defense in mission-critical environments requires integrated approaches capable of representing adversarial progression, defender-side uncertainty, mission impact, and defensive decision support within a unified framework. In operational domains, defenders must continuously estimate the evolving security posture while preserving the continuity and integrity of mission-critical functions under incomplete and noisy observations. This paper presents a mission-oriented cyber-defense framework for Cyber Situational Awareness (CSA) and decision support based on a three-layer integrated probabilistic model and an executable simulation prototype. The model combines: (i) an attack-graph model that represents possible adversarial progression through mission-relevant assets, (ii) an event model that transforms observed telemetry into posterior defender beliefs through Bayesian inference, and (iii) a state model that abstracts the inferred posture into conflict states and mission-risk levels. These components are connected to a one-step defensive action rule that balances estimated residual mission risk and operational cost. The framework is instantiated in a NetLogo agent-based simulation of an operational environment structured across the Tactical Edge Zone (TEZ), Mission Operations Zone (MOZ), and Enterprise Support Zone (ESZ). The proposal is assessed through mathematical consistency analysis, local robustness assessment under telemetry perturbations, and representative simulation traces. Results indicate that the framework and its implementation preserve coherent relationships between attack progression, telemetry-driven uncertainty management, mission-impact assessment, and cost-aware defensive decision support guided by mission-risk prioritization.
\end{abstract}


\section{Introduction}\label{sec:introduction}

Cyber defense in operational environments requires more than identifying isolated vulnerabilities or reacting to individual alerts. In mission-critical settings, defenders must understand how adversaries may progress across interconnected assets, interpret partial and noisy telemetry, and select defensive actions without compromising mission continuity. Within this mission-oriented perspective, technical compromise becomes relevant when it affects the functions and tasks that sustain operations, e.g. maintaining Intelligence, Surveillance, and Reconnaissance (ISR) availability, or preserving Command and Control (C2) continuity. This operational context allows defenders to assess risk, interpret escalation, and prioritize responses according to their potential mission impact. Importantly, cyber defense must distinguish between technical compromise and operational impact: the compromise of several Tactical Edge Zone (TEZ) nodes may represent a significant warning without yet affecting the mission, whereas the compromise of a single mission-critical node may produce a disproportionate increase in operational mission-risk. 

A central difficulty is that the defender rarely observes the real adversarial state directly. Attacker progression is hidden, monitoring coverage is incomplete, and observable events provide only indirect evidence about what may be happening inside the system. At the same time, defensive decisions cannot wait for perfect information. The defender must maintain an evolving estimate of the current posture, evaluate mission impact, and select actions that reduce risk under operational constraints. 

Existing cyber defense representations each capture only part of the problem, thereby leading to a partial Cyber Situational Awareness (CSA): multi-step adversarial progression models provide structural visibility but do not fully integrate live evidence updates; event- and alert-based models reflect telemetry but require probabilistic linkage to underlying attack activity; and decision-theoretic models address partial observability but are often hard to implement in cyber defense scenarios. This gap motivates more compact frameworks for CSA that unify adversarial progression, telemetry inference, operational state abstraction, mission risk, and defensive response within a single coherent loop.

Considering the previous challenges, this paper proposes an integrated mission-oriented cyber-defense framework for CSA that combines a three-layer mathematical model with an executable simulation prototype. Together, these layers capture hidden attacker dynamics, update probabilistic beliefs from noisy observations, summarize inferred system posture, and link mission-level risk to defensive action selection through a one-step decision rule. This structure is operationalized in a NetLogo-based simulation, a programmable agent-based modeling environment for complex evolving systems~\cite{NetLogo}. Thus, the main contributions of this paper are threefold:
\begin{enumerate}[label=\roman*)]
    \item \textbf{A three-layer mathematical mission-oriented cyber-defense model.} The proposal integrates a graph model for adversarial progression, an event model for telemetry-driven Bayesian inference, and a state model for conflict-state abstraction and mission-risk assessment.

    \item \textbf{Analytical and scenario-based validation.} The model is assessed through mathematical consistency arguments, local robustness analysis under perturbations of the observation model, and representative simulation traces.

    \item \textbf{An executable NetLogo prototype.} The mathematical model is adapted into an interactive simulation of a scenario, providing an experimental environment for inspecting how attacker progression, mission impact, and defensive response interact over time. The simulator code, scenario files, plotting scripts, and representative execution traces are available in a companion repository.\footnote{\url{https://github.com/requenams/mission-oriented-cyber-defense-netlogo}}
\end{enumerate}

The remainder of the paper is organized as follows. Section~\ref{sec:related_work} reviews related work. Section~\ref{sec:mathematical_model} introduces the proposed three-layer mathematical model. Section~\ref{sec:simulation_adaptation} describes the simulation-oriented adaptation. Section~\ref{sec:results_validation} presents the validation results. Section~\ref{sec:conclusions} concludes the paper and outlines future research directions.
\section{Related Work}
\label{sec:related_work}

Cyber-defense scenarios can be represented from several complementary perspectives. Some approaches focus on structural attack paths, others incorporate probabilistic reasoning from evidence, others model temporal evolution and decision-making under uncertainty, and others provide standardized vocabularies for describing adversary behavior and defensive actions. This section groups the most relevant literature into the following families: attack graphs and attack-defense trees, Bayesian Attack Graphs, Markovian and partially observable decision models, fuzzy and game-theoretic models. These approaches contribute in different ways to the levels of CSA, particularly to the comprehension level.

Attack graphs are a central formalism for representing multi-step adversarial progression. Early graph-based approaches showed how vulnerabilities, reachability, and privilege dependencies can be combined to analyze attack paths~\cite{phillips1998graphbased}. Logic-based systems such as MulVAL encode preconditions and postconditions through rules~\cite{ou2005mulval}, while practical tools such as NetSPA emphasize scalable generation for network defense~\cite{NetSPA}. Surveys and taxonomies highlight the value of attack graphs for structural reasoning, but also point out their limitations when uncertainty and live evidence must be incorporated~\cite{MulVALSurvey}. Attack-defense tree models and related extensions include countermeasures~\cite{ADT_JLC}, supporting the representation of attacker goals and defensive responses. However, tree-based representations are less natural for graph-like lateral movement across operational zones and do not provide telemetry-driven belief updates.

Bayesian Attack Graphs address part of this limitation by combining attack graph structure with Bayesian-network semantics. Frigault et al. introduced Bayesian-network-based approaches for measuring network security and dynamic attack progression~\cite{frigault2008,frigault2010}. Poolsappasit et al. proposed dynamic security risk management using Bayesian Attack Graphs~\cite{BAG}, while Mu{\~n}oz-Gonz{\'a}lez et al. studied exact inference techniques for this family of models~\cite{munoz2019}. Other works use Bayesian networks for real-time alert correlation and prediction~\cite{BAG3}. Moreover, mission-centric extensions are particularly relevant: Javorn\'ik et al.~\cite{javornik2019mission} and Javorn\'ik and Hus\'ak~\cite{BAG2} combine attack-graph representations, Bayesian reasoning, and mission-oriented decision support. These approaches explicitly consider the mission impact, but they do not fully integrate adversarial progression, telemetry-driven belief updates, mission-state abstraction, and executable simulation within a single CSA loop.

Temporal probabilistic models are also relevant because intrusions unfold over time and defenders must reason from partial observations. Hidden Markov Models have been used for alert correlation, APT prediction, and cyber-physical intrusion detection~\cite{HMM1,zhang2022msahmm}. They are useful when attacks can be represented as sequences of hidden stages emitting observations, but their sequential structure may compress branching attacker behavior. POMDPs provide a principled framework for acting under partial observability~\cite{monahan1982state,kaelbling1998planning}. In cyber defense, they have been used for dynamic defense of large-scale networks~\cite{POMDP2}, adaptive defense against multi-stage attacks~\cite{POMDP1}, and joint defense and monitoring under uncertainty~\cite{POMDP3}. Nevertheless, full POMDP solutions often require detailed transition, observation, and reward models and may be computationally demanding. This motivates lighter decision-support mechanisms that preserve the belief-state intuition while remaining easier to instantiate in concrete cyber-defense scenarios. 

Fuzzy Cognitive Maps provide another way to represent cybersecurity scenarios through weighted causal relations and qualitative risk propagation~\cite{FCM,felix2019fcm_review}. Their main strength is interpretability, especially when expert knowledge is used to define causal dependencies between system variables. However, standard FCMs do not naturally provide Bayesian belief updates from telemetry over multi-step attack dependencies. Game-theoretic models capture strategic attacker-defender interaction and adaptive defense~\cite{GTSurveyAPT}. They are useful for analyzing adversarial incentives, equilibrium behavior, and moving-target defense, but their focus differs from defender-centric situational awareness based on partial telemetry, mission-risk assessment, and immediate response selection.

Motzek et al.~\cite{motzek2017pareto} compute Pareto-efficient response plans by balancing cyber risk and operational impact, highlighting that mitigation actions may reduce risk while degrading mission performance. Similar trade-off formulations are also explored in game-theoretic cyber defense models and POMDP-based decision frameworks. In contrast, this work adopts a simpler one-step decision rule that combines residual mission risk and operational cost, avoiding full Pareto-frontier computation while remaining suitable for a CSA prototype.

Overall, existing approaches provide valuable yet often fragmented capabilities. Attack graphs offer structural clarity regarding adversarial progression; Bayesian attack graphs introduce probabilistic inference; Markovian and POMDP-based models address temporal uncertainty and decision-making; FCMs and game-theoretic models support causal and strategic reasoning; and standardized knowledge bases improve the vocabulary used to describe attacker and defender behavior. Each of these mathematical approaches contributes to understanding a specific element of the scenario. However, current research still lacks sufficient integration across models to enable a holistic view of the full cyber defense scenario, spanning from atomic-level technical events to broader, mission-oriented considerations. Such integration could contribute to the maturity level of CSA.

\section{Mathematical Mission-Oriented Cyber-Defense Model}
\label{sec:mathematical_model}

This section introduces the three-layer integrated mathematical probabilistic model of the proposed CSA framework. The objective is to connect three complementary views of the cyber-defense problem: a graph model for adversarial progression, an event model for telemetry-driven inference, and a state model for operational abstraction and mission risk. These components are then combined with a defensive action rule.

The model is organized around a directed progression graph \(G\), a set of hidden adversarial variables \(X\), a telemetry event set \(E\), a conflict-state space \(S\), a defensive action space \(A\), a conflict-state mapping \(\Psi\), and a mission-risk functional \(R\). The graph model describes how adversarial conditions enable one another, the event model updates the defender's belief from observed telemetry, and the state model translates this belief into operational posture and mission-oriented risk. 

From the defender's perspective, this organization makes uncertainty management an explicit part of CSA: the true adversarial configuration is treated as hidden, telemetry provides partial and noisy evidence, and Bayesian belief updates transform this evidence into posterior compromise probabilities. These posterior estimates then support both mission-risk assessment and decision support, since defensive actions are selected by balancing expected residual mission risk against operational cost.

The application of these models can be related to the phases of Situational Awareness (SA) defined by Endsley~\cite{EndsleySA}: Perception using telemetry and events; Comprehension with telemetry and belief-update; Projection regarding graphs, predictive belief, and mission-risk; and Decision through the selection of defensive actions.

\subsection{Graph model: adversarial progression}
\label{subsec:graph_model}

The graph model represents how an attacker may progress through the defended system. Let $G=(V,\mathcal{R})$ be a directed acyclic progression graph, where $V={v_1,\ldots,v_n}$ is the set of adversarial progression nodes and $\mathcal{R}\subseteq V\times V$ is the set of directed enabling relations. Hidden adversarial progression is encoded by binary random variables
\begin{equation}
    X=\{X_1,\ldots,X_n\}, 
    \qquad 
    X_i\in\{0,1\},
\end{equation}
where each variable \(X_i\) is associated with a node or step in the progression graph. From a cyber-defense perspective, \(X_i=1\) means that the corresponding attack condition, procedure step, or capability has been reached or activated by the adversary, while \(X_i=0\) means that it has not. A concrete realization of the hidden progression is denoted by
\begin{equation}
    x=(x_1,\ldots,x_n)\in\Omega,
    \qquad
    \Omega=\{0,1\}^n
\end{equation}
Thus, \(X\) denotes the random adversarial progression, while \(x\) denotes one specific attack configuration: a precise assignment indicating which graph nodes have or have not been reached by the adversary.

The progression graph \(G=(V,\mathcal{R})\) is assumed to be directed and acyclic, with nodes associated with variables in \(X\). Edges represent prerequisite or enabling relations between adversarial conditions. If \(Pa(X_i)\) denotes the parent set of \(X_i\), the local conditional probability
\begin{equation}
    \theta_i(u)=P(X_i=1\mid Pa(X_i)=u)
\end{equation}
represents the probability that node \(X_i\) becomes active, given that its parent nodes are active or inactive according to the assignment \(u\). In operational terms, this captures how previous attacker progress enables or increases the likelihood of reaching a subsequent procedure step. The probability of a complete attack configuration $x$ is obtained by combining all local conditional probabilities:
\begin{equation}
P(x)=
\prod_{i=1}^{n}
\theta_i(x_{Pa(X_i)})^{x_i}
\left(1-\theta_i(x_{Pa(X_i)})\right)^{1-x_i}
\label{eq:joint_factorization}
\end{equation}
Therefore, $P(x)$ gives the probability of a full adversarial state, that is, the probability that the attacker has reached exactly the set of graph nodes indicated by $x$. Thus, the graph model captures adversarial progression as a structured probabilistic dependency model rather than as an unstructured collection of alerts.

\subsection{Event model: telemetry and belief update}
\label{subsec:event_model}

The defender does not observe the hidden state \(x\) directly. Instead, at time \(t\), it observes a telemetry batch \(O_t\subseteq E\), where \(E=\{E_1,\ldots,E_m\}\). In a cyber-defense scenario, these events represent alerts, sensor outputs, anomalous behaviors, or other observable evidence generated by the monitoring infrastructure.

Each event \(E_k\) is associated with a hidden adversarial variable \(X_{\nu(k)}\) through a mapping
\begin{equation}
    \nu:\{1,\ldots,m\}\rightarrow\{1,\ldots,n\}.
\end{equation}
This means that event \(E_k\) is interpreted as evidence about whether the corresponding attack step \(X_{\nu(k)}\) has been reached. For each event, the observation parameters are
\begin{equation}
    \lambda_k^1=P(E_k=1\mid X_{\nu(k)}=1),
    \qquad
    \lambda_k^0=P(E_k=1\mid X_{\nu(k)}=0),
\end{equation}
with \(\lambda_k^1>\lambda_k^0\) in typical cases. Thus, an event is more likely to be observed when the monitored adversarial condition is active, while still allowing false positives or noisy observations.

Under a positive-evidence model,
\begin{equation}
    P(O_t\mid x)=
    \prod_{E_k\in O_t}
    P(E_k=1\mid x),
    \qquad
    P(E_k=1\mid x)=
    \begin{cases}
        \lambda_k^1, & x_{\nu(k)}=1,\\
        \lambda_k^0, & x_{\nu(k)}=0.
    \end{cases}
    \label{eq:telemetry_likelihood}
\end{equation}
This likelihood quantifies how compatible the observed telemetry batch is with a given attack configuration. If absence of monitored events is informative, the likelihood can be extended with the corresponding factors for \(E_k\in E\setminus O_t\).

Let \(b_t(x)\) denote the defender's belief after processing observations up to time \(t\). Given a predictive belief \(b_{t^-}\), the Bayesian update is
\begin{equation}
    b_t(x)=
    \frac{
        b_{t^-}(x)P(O_t\mid x)
    }{
        \sum_{x'\in\Omega}b_{t^-}(x')P(O_t\mid x')
    }.
    \label{eq:bayesian_update}
\end{equation}
This update transforms observed events into an updated probability distribution over possible adversarial configurations.

The posterior marginals
\begin{equation}
    p_i^{(t)}
    =
    P(X_i=1\mid O_{1:t})
    =
    \sum_{x\in\Omega}x_i b_t(x)
    \label{eq:marginals}
\end{equation}
summarize the inferred probability that each graph node or attack step has been reached. These probabilities provide a compact view of the defender's current estimate without requiring direct inspection of the full hidden state space.

\subsection{State model: conflict abstraction and mission risk}
\label{subsec:state_model}

The state model maps posterior marginals and observations into a compact conflict-state space
\begin{equation}
    S=\{S_0,S_1,S_2,S_3,S_4,S_5\},
\end{equation}
representing increasing severity from normal conditions to mission impact.

For a group \(G_c\subseteq\{1,\ldots,n\}\) associated with an operational phase, define the compromise aggregator
\begin{equation}
    r_{G_c}(p)=1-\prod_{i\in G_c}(1-p_i).
    \label{eq:or_aggregator}
\end{equation}
This quantity summarizes the probability that at least one relevant attack step in the group is active. Let \(r_{TEZ}\), \(r_{PIV}\), and \(r_{MIS}\) denote the aggregators for tactical-edge compromise, pivot/lateral movement, and mission-impact capability.

Given thresholds \(\tau_1,\ldots,\tau_4\) and a set \(I\subseteq E\) of direct mission-impact events, conflict states are assigned by priority:
\begin{equation}
    \Psi(p,O_t)=
    \begin{cases}
        S_5, & O_t\cap I\neq\emptyset,\\
        S_4, & r_{MIS}\geq\tau_4,\\
        S_3, & r_{PIV}\geq\tau_3,\\
        S_2, & r_{TEZ}\geq\tau_2,\\
        S_1, & r_{TEZ}\geq\tau_1,\\
        S_0, & \text{otherwise.}
    \end{cases}
    \label{eq:conflict_mapping}
\end{equation}
The rule prioritizes direct mission-impact evidence, followed by mission-risk, pivot, and tactical-edge indicators. In operational terms, this converts probabilistic inference into a discrete posture that can be more easily interpreted by a defender. Mission risk connects cyber compromise with operational impact. In the selected scenario, the attacker is assumed to pursue degradation of ISR, disruption of C2, and compromise of COP/RMP integrity. From the defender's perspective, these define the main mission-critical functions. Let \(p_{COP}\), \(p_{ISR}\), and \(p_{C2}\) denote the corresponding posterior compromise probabilities. With nonnegative weights \(w_{COP},w_{ISR},w_{C2}\), the mission-risk functional is
\begin{equation}
    R(b_t)=
    w_{COP}p_{COP}^{(t)}
    +
    w_{ISR}p_{ISR}^{(t)}
    +
    w_{C2}p_{C2}^{(t)}.
    \label{eq:mission_risk}
\end{equation}
This score is bounded by \(w_{COP}+w_{ISR}+w_{C2}\) and increases with the inferred compromise probability of mission-critical functions. It therefore distinguishes between technical compromise and mission-level impact.

\subsection{Model integration and defensive actions}
\label{subsec:model_integration_actions}

\begin{figure}[]
    \centering
    \includegraphics[width=0.74\linewidth]{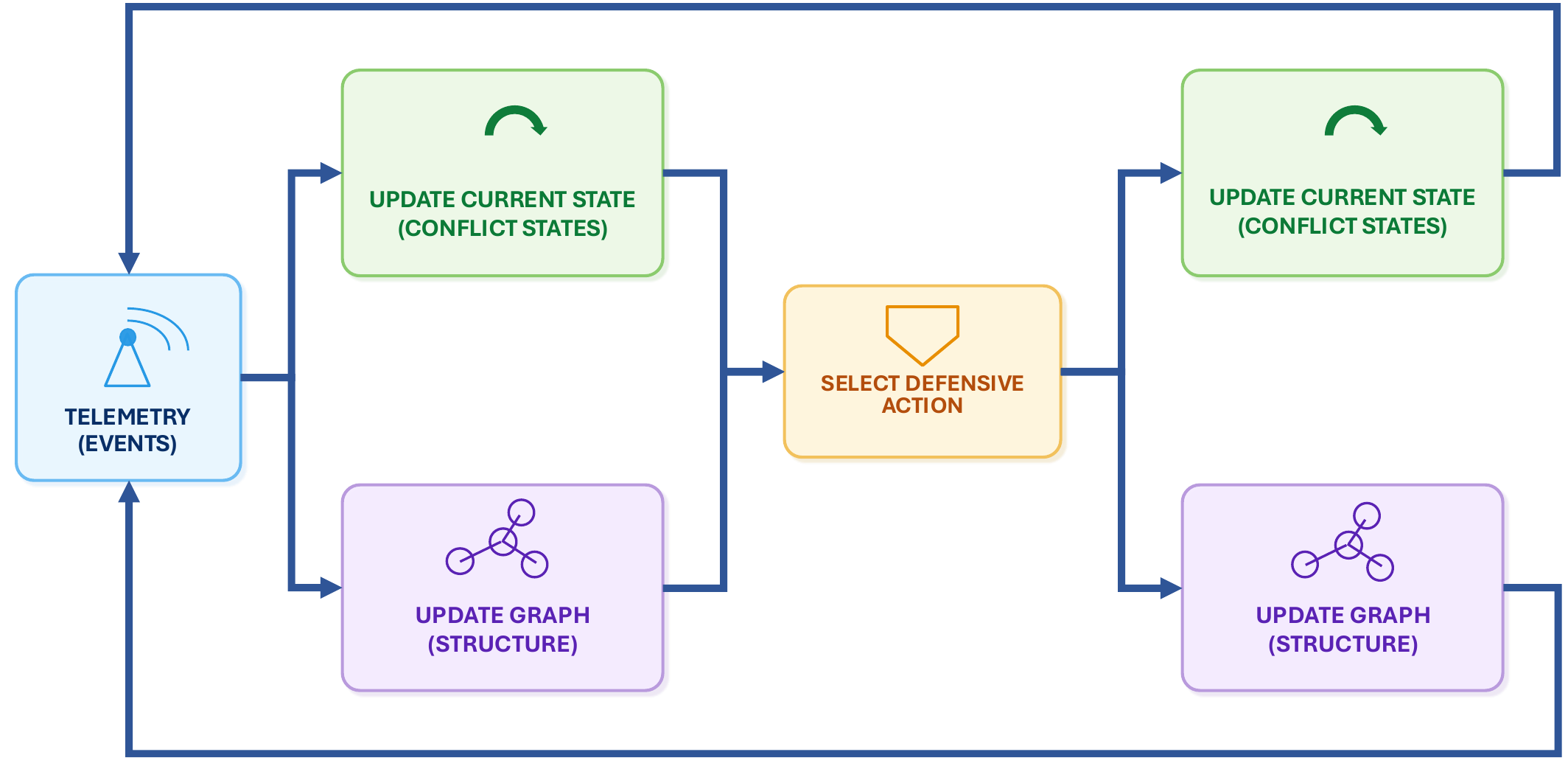}
    \caption{Integration of the graph, event, and state models into the mission-oriented cyber-defense loop.}
    \label{fig:three_model_integration}
\end{figure}

The three models are combined into a defender-centric decision loop. The graph model defines possible adversarial progression, the event model updates the defender's belief from telemetry, and the state model converts posterior information into operational posture and mission risk. Figure~\ref{fig:three_model_integration} summarizes this interaction.

The framework selects defensive actions through a one-step decision rule. Let \(A\) be a finite set of possible defensive actions and let \(A(s)\subseteq A\) be the subset of actions admissible in conflict state \(s\). These actions may include, for example, monitoring, telemetry improvement, hardening, quarantine, path blocking, containment, or protection of mission-critical assets, but the mathematical definition is not restricted to this specific list.

For each candidate action \(a\in A(s_t)\), the defender estimates a post-action risk \(R_a(b_t)\) and an operational cost \(C_a\geq 0\). The post-action risk represents the estimated residual mission risk after applying action \(a\) from the current belief \(b_t\). The operational cost represents the expected burden of applying the action, such as loss of availability, resource consumption, disruption of normal operations, or operator workload. The selected action is
\begin{equation}
    a_t^\star
    \in
    \arg\min_{a\in A(s_t)}
    \left(R_a(b_t)+C_a\right),
    \qquad
    s_t=\Psi(p^{(t)},O_t).
    \label{eq:one_step_decision}
\end{equation}
This rule does not claim long-horizon optimality. Instead, it provides a tractable immediate-response mechanism that balances estimated mission-risk reduction against operational cost. The resulting pipeline can be read as follows: observed alerts \(O_t\) update the defender's belief \(b_t\); this belief is used to compute posterior compromise probabilities \(p^{(t)}\); these probabilities, together with the current observations \(O_t\), determine the conflict state \(s_t\); finally, the conflict state and the mission risk \(R(b_t)\) guide the selection of the immediate defensive action \(a_t^\star\):
\begin{equation}
    O_t
    \longrightarrow
    b_t
    \longrightarrow
    p^{(t)}
    \longrightarrow
    s_t
    \longrightarrow
    R(b_t)
    \longrightarrow
    a_t^\star.
    \label{eq:mathematical_pipeline}
\end{equation}

\subsection{Mathematical consistency and local robustness}
\label{subsec:mathematical_consistency_robustness}

The mathematical model satisfies several basic consistency properties. The Bayesian update is well defined whenever the normalization constant is strictly positive, in which case \(b_t\) is a valid probability distribution over the finite hidden state space. Consequently, the marginals \(p_i^{(t)}\) lie in \([0,1]\). The OR-style aggregators are bounded and monotone, the conflict-state map returns a unique state for every fixed pair \((p,O_t)\), and the mission-risk functional is finite, nonnegative, and bounded by the sum of its mission weights. Since the admissible action set is finite, the one-step decision rule also admits at least one minimizing action.

The model was also analyzed under local perturbations of the observation likelihood. Let \(L_t(x)=P(O_t\mid x)\) be the nominal likelihood and let \(\widetilde{L}_t(x)\) be a perturbed likelihood such that
\[
    \sup_{x\in\Omega}|L_t(x)-\widetilde{L}_t(x)|\leq \varepsilon .
\]
When the Bayesian normalization constant remains bounded away from zero, small perturbations of \(L_t\) induce bounded changes in the posterior belief. Since marginals, aggregate scores, and mission risk are deterministic functions of the posterior, their variations are also controlled. Moreover, if aggregate scores remain separated from conflict-state thresholds, and if the selected action has a positive objective margin over the alternatives, then the discrete outputs---conflict state and selected action---remain unchanged under sufficiently small perturbations. This provides a local robustness argument for settings where telemetry parameters are uncertain but not arbitrary.
\section{Simulation-Oriented Adaptation and Practical Implementation}
\label{sec:simulation_adaptation}\label{subsec:mathematical_to_sim_mapping}

This section describes how the framework is adapted into an executable NetLogo prototype. The goal is to build an interpretable simulation that preserves the main structure of the proposal while enhancing CSA and decision-support: staged adversarial progression, partial telemetry, conflict-state abstraction, mission-risk assessment, and defensive response. The implementation represents adversarial progression through explicit graph nodes. This reduces abstraction but increases interpretability: the user can directly observe compromised, protected, quarantined, and exposed assets at each simulation tick. The mathematical adapted components are explained in the following subsections, and are also summarized in Table~\ref{tab:mathematical_sim_mapping}.

\begin{table}[h]
    \centering
    \caption{Main correspondence between the mathematical model and the NetLogo prototype.}
    \label{tab:mathematical_sim_mapping}
    \footnotesize
    \begin{tabular}{p{0.34\linewidth}p{0.56\linewidth}}
        \hline
        Mathematical component & Simulation-oriented instantiation \\
        \hline
        Hidden adversarial variables and progression graph
        & Explicit propagation graph with 19 nodes: 6 in TEZ, 8 in MOZ, and 5 in ESZ, connected by directed propagation links. \\

        Telemetry events and partial observation
        & Generated and observed event families \(E_1,\ldots,E_7\), including weak \texttt{PRE@Node} alerts. \\

        Conflict-state abstraction
        & Classification of six-level postures based on mission-risk and graph status: \(S0,\ldots,S5\) \\
   
        Mission-risk functional
        & Weighted score over compromised assets, with higher weights for \texttt{COP}, \texttt{ISR}, and \texttt{C2}. \\

        Defensive action rule
        & Manual actions or automatic action-target recommendations based on the current state, mission risk, telemetry, target relevance, and admissible actions. \(A0,\ldots,A6\) \\ \\
        \hline
    \end{tabular}
\end{table}

The simulator exposes explicit node states to improve interpretability and interactive inspection while preserving the conceptual structure of the mathematical framework. Furthermore, although the simulator does not implement a formal ATT\&CK~\footnote{\url{https://attack.mitre.org}}/D3FEND~\footnote{\url{https://d3fend.mitre.org}} mapping, the staged adversarial progression, telemetry categories, and defensive actions are conceptually aligned with the type of attacker behaviors, observable procedures, and countermeasure abstractions commonly represented in MITRE knowledge bases. 

\subsection{Node-based scenario}
\label{subsec:from_hidden_to_nodes}

The simulated environment is a directed graph
\begin{equation}
    \mathcal{G}_{\mathrm{sim}}=(N,L),
\end{equation}
where \(N\) is the set of simulated assets and \(L\subseteq N\times N\) represents admissible propagation paths. Nodes are distributed across three zones: the Tactical Edge Zone (TEZ), the Mission Operations Zone (MOZ), and the Enterprise Support Zone (ESZ). TEZ represents tactical-edge communications and entry points, MOZ contains mission-operation assets, and ESZ represents enterprise-support services.

Each node \(v\) has a zone \(z(v)\), a functional type \(\kappa(v)\), and dynamic state variables. For example, \(c_t(v)\) is binary and indicates whether node \(v\) is compromised at time \(t\): if \(c_t(v)=1\), the node is compromised; if \(c_t(v)=0\), it is not. Other node-level variables represent protection or hardening, quarantine effects, detection capability, and telemetry quality.

The implemented scenario includes TEZ nodes such as \texttt{TEZ-GW}, \texttt{SATCOM}, \texttt{RADIO}, \texttt{EDGE}, and \texttt{TEZ-B1/B2}; MOZ nodes such as \texttt{MOZ-IN}, \texttt{RELAY}, \texttt{OPS-WS}, \texttt{MOZ-ADM}, \texttt{MOZ-B}, \texttt{COP}, \texttt{ISR}, and \texttt{C2}; and ESZ nodes such as \texttt{ESZ-IN}, \texttt{IDP}, \texttt{SIEM}, \texttt{REPO}, and \texttt{HQ}. Here, \texttt{SATCOM} denotes satellite communications, \texttt{COP} the Common Operational Picture, \texttt{ISR} Intelligence, Surveillance, and Reconnaissance, \texttt{C2} Command and Control, \texttt{IDP} an identity provider, \texttt{SIEM} a security information and event management system, \texttt{REPO} a repository, and \texttt{HQ} headquarters support (see graph in Figure~\ref{fig:scenario_initial}).

\subsection{Attack propagation and telemetry}
\label{subsec:attack_propagation}

Attack progression is implemented as a discrete-time propagation process over the directed scenario graph. At each tick, compromised nodes may influence reachable non-compromised nodes according to zone-dependent propagation parameters, node-specific risk factors, and the current defensive configuration. Defensive actions can reduce propagation by hardening nodes, quarantining compromised assets, or restricting selected outgoing paths.

This mechanism is intended as a simulation-oriented counterpart of the progression graph. The mathematical model represents adversarial progression through hidden variables and probabilistic dependencies, whereas the simulator represents progression through explicit node compromise. Although this is a practical simplification, both representations preserve the same operational intuition: early compromise can enable later compromise, and defensive actions can modify the likelihood or feasibility of subsequent progression.

Telemetry is generated from the simulated compromise process. The prototype uses a compact event vocabulary \(E_1,\ldots,E_7\). These event families correspond to broad operational categories: \texttt{E1} for TEZ entry activity, \texttt{E2} for internal TEZ activity, \texttt{E3} for MOZ entry activity, \texttt{E4} for internal MOZ movement, \texttt{E5} for mission-critical compromise, \texttt{E6} for ESZ entry activity, and \texttt{E7} for internal ESZ activity. In addition, the simulator includes weak pre-alerts of the form \texttt{PRE@Node}, which represent suspicious evidence around exposed but not yet compromised nodes.

The observation process is intentionally partial. Events may be generated internally without necessarily being observed by the defender, depending on the current monitoring and telemetry configuration. This preserves the partial-observability principle of the model while keeping the implementation simple enough for interactive simulation.

\subsection{Conflict states, mission risk, and actions}
\label{subsec:sim_state_risk}

The simulator preserves the six-state abstraction introduced in the framework: \(S=\{S_0,S_1,S_2,S_3,S_4,S_5\}\). These states correspond to \texttt{S0-Normal}, where no relevant compromise has been detected; \texttt{S1-Initial-Compromise}, where the attack is still limited to the tactical edge; \texttt{S2-TEZ-Frontier}, where frontier nodes near the transition to MOZ are affected; \texttt{S3-MOZ-Lateral-Movement}, where the adversary has entered the mission-operation zone; \texttt{S4-Mission-Risk}, where mission-support or boundary assets are threatened; and \texttt{S5-Mission-Impact}, where mission-critical assets or substantial ESZ compromise are affected. In the prototype, conflict states are approximated through operational rules over the explicit compromise configuration. Therefore, the state escalation is directly inspectable during the execution.

Mission risk is implemented as a weighted score over compromised assets. Higher weights are assigned to assets that are directly connected to the mission objectives of the scenario, especially \texttt{COP}, \texttt{ISR}, and \texttt{C2}. Thus, the risk score is not merely a count of compromised nodes: it reflects the operational relevance of the affected assets. This keeps the practical implementation aligned with the mission-risk function of the model.

The executable action space is \(A_{\mathrm{sim}}=\{A_0,A_1,A_2,A_3,A_4,A_5,A_6\}\), where \texttt{A0-Monitor} observes the current state, \texttt{A1-Quarantine-Node} reduces propagation, \texttt{A2-Harden-Neighbors} protects outgoing neighbor nodes, \texttt{A3-Boost-Telemetry} improves detection, \texttt{A4-Block-Outgoing} restricts outgoing propagation paths, \texttt{A5-Protect-Critical} reinforces mission-critical assets, and \texttt{A6-Contain-TEZ} applies early containment in TEZ. Together, these actions provide the core defensive capabilities required by the scenario: visibility, containment, hardening, segmentation, and protection of mission-critical assets.

\subsection{Manual and automatic execution modes}
\label{subsec:sim_decision_modes}
The simulator supports two execution modes. In manual mode, the operator selects both the defensive action and the target node. This mode is intended for exploratory analysis, allowing different defensive strategies to be tested under the same scenario structure. In automatic mode, the simulator recommends an action-target pair at each tick. This recommendation mechanism is an implementation-oriented approximation of the one-step decision rule introduced in Section~\ref{subsec:model_integration_actions}. Instead of solving the optimization problem exactly, the prototype evaluates admissible actions according to the current conflict state, the mission-risk score, the observed telemetry, the importance of the candidate target, its compromise status, and the expected local effect of the action.

For each conflict state, only a subset of actions is admissible. For example, early states prioritize monitoring, telemetry improvement, and TEZ containment, whereas later states enable quarantine, outgoing-path blocking, and protection of mission-critical assets. This action gating ensures that the automatic policy remains consistent with the operational meaning of the conflict states.

In practical terms, each simulation tick follows the same operational loop. The current node-compromise configuration, observed telemetry, conflict state, and mission-risk score determine the defensive response. The selected action and target are then applied, the attack propagates through the graph, new telemetry is generated and filtered, and the conflict state and mission-risk score are updated for the next tick:
\begin{equation}
    (c_t,O_t,s_t,R_t^{\mathrm{sim}})
    \longrightarrow
    (a_t,v_t)
    \longrightarrow
    c_{t+1}
    \longrightarrow
    O_{t+1}
    \longrightarrow
    (s_{t+1},R_{t+1}^{\mathrm{sim}}).
    \label{eq:sim_pipeline}
\end{equation}

A final implementation decision is the separation between simulation logic and scenario data. Nodes, links, propagation values, and mission-risk weights are loaded from external CSV files rather than hardcoded in the NetLogo source code. The simulator also exports textual logs, probability-calculation traces, and node-level snapshots, which makes the representative executions auditable and supports the validation presented in Section~\ref{sec:results_validation}.
\section{Results and Scenario Validation}
\label{sec:results_validation}
This section evaluates the NetLogo prototype through a representative execution of a cyber defense scenario. The objective is not to claim operational validation against real incidents, but to assess whether the implemented simulation behaves coherently under controlled conditions. In particular, the validation checks whether attacker progression, telemetry generation, conflict-state escalation, mission-risk evolution, and defensive actions remain aligned with the mission-oriented structure of the proposed framework.

The simulation layer was evaluated through an automatic execution exported from the NetLogo prototype, where the best recommended defense action is always selected. The scenario starts from an initial compromise in the Tactical Edge Zone and evolves through the TEZ--MOZ--ESZ structure. The attacker's mission-relevant objectives are associated with degradation or compromise of ISR, C2, and COP/RMP; therefore, the defender's most critical assets are \texttt{ISR}, \texttt{C2}, and \texttt{COP}. The reported milestones and plots are derived from exported simulation traces. The simulator code, scenario CSV files, plotting scripts, and representative execution logs are available in the companion repository\footnote{\url{https://github.com/requenams/mission-oriented-cyber-defense-netlogo}}.

\subsection{Simulation setup and visual inspection}
\label{subsec:simulation_visual_results}

\begin{figure}[h]
    \centering
    \includegraphics[width=\linewidth]{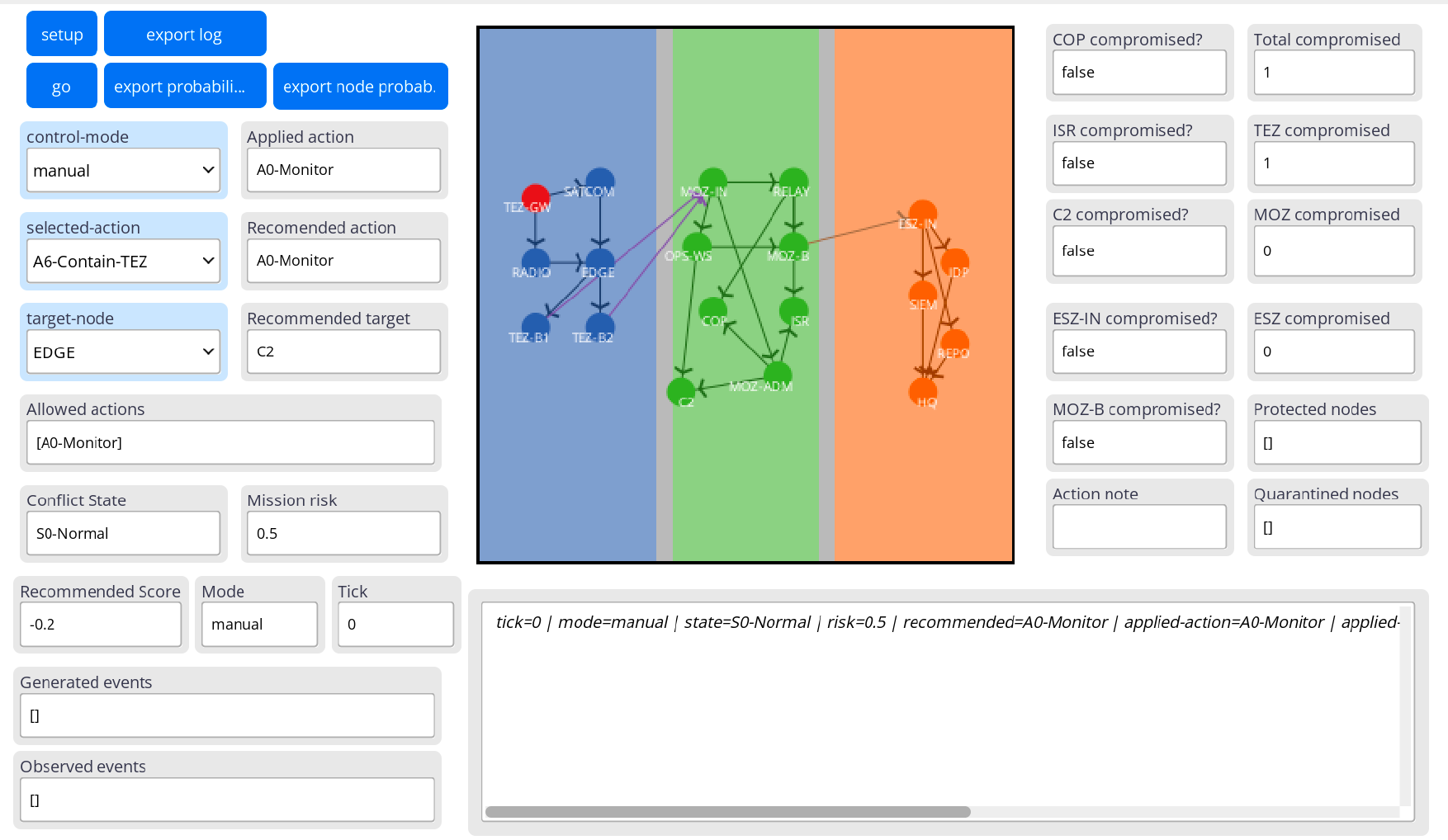}
    \caption{NetLogo interface for the representative automatic execution at tick 0.}
    \label{fig:scenario_initial}
\end{figure}

The initial configuration of a representative execution can be seen in
\figurename{\ref{fig:scenario_initial}}. The view includes the TEZ--MOZ--ESZ topology, the available manual and automatic controls, the mission-critical status indicators, the conflict-state and mission-risk monitors, the execution plots, and the initial log entry. At this point, the scenario is in the initial low-risk condition, with only the forced \texttt{TEZ-GW} compromise represented in the TEZ zone. 

\subsection{Representative automatic execution}
\label{subsec:automatic_results}
\begin{figure}[h]
    \centering
    \begin{subfigure}{0.48\linewidth}
        \centering
        \includegraphics[width=\linewidth]{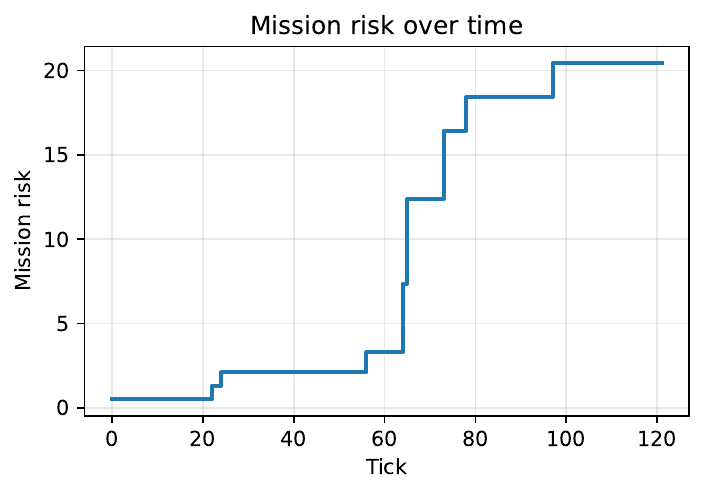}
        \caption{Mission risk.}
        \label{fig:auto_mission_risk}
    \end{subfigure}
    \begin{subfigure}{0.48\linewidth}
        \centering
        \includegraphics[width=\linewidth]{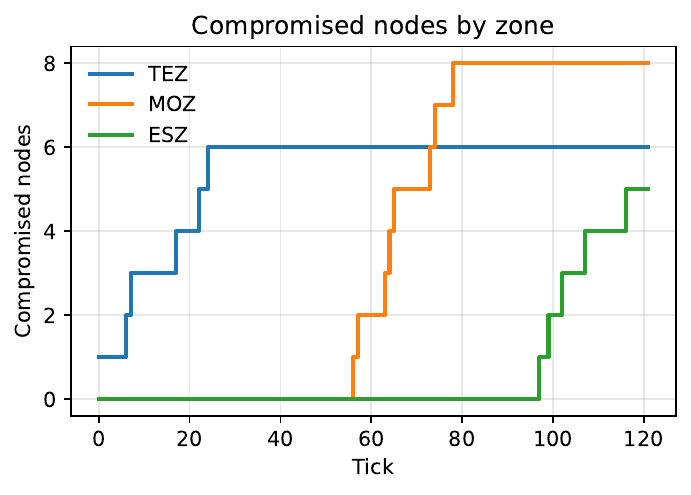}
        \caption{Compromised nodes by zone.}
        \label{fig:auto_compromised_by_zone}
    \end{subfigure}
    \vspace{0.2em}
    \begin{subfigure}{0.48\linewidth}
        \centering
        \includegraphics[width=\linewidth]{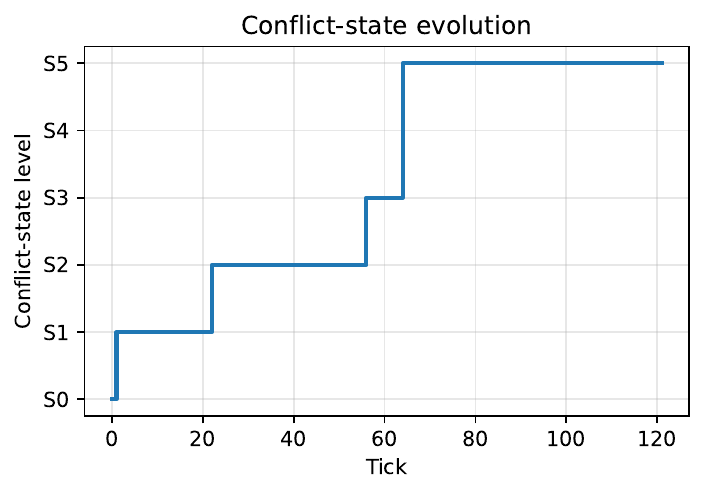}
        \caption{Conflict-state evolution.}
        \label{fig:auto_conflict_state}
    \end{subfigure}
    \begin{subfigure}{0.48\linewidth}
        \centering
        \includegraphics[width=\linewidth]{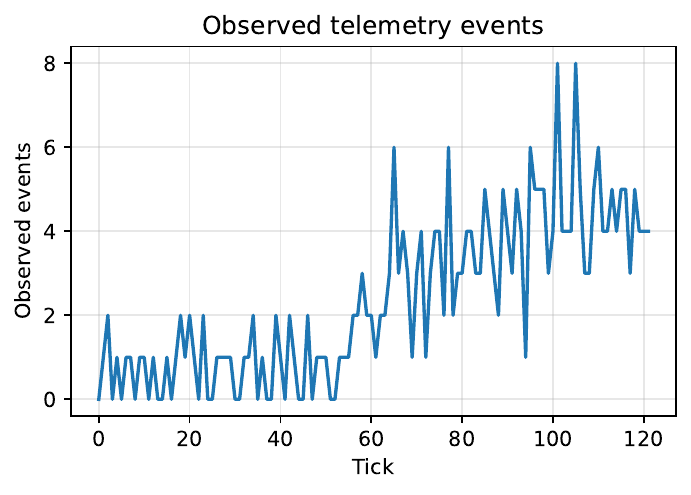}
        \caption{Observed telemetry events.}
        \label{fig:auto_observed_events}
    \end{subfigure}

    \caption{Evolution of mission risk, compromised nodes, conflict state, and observed telemetry during the representative automatic mode execution (120 ticks).}
    \label{fig:auto_plots}
\end{figure}

The results obtained from running the simulation in automatic mode are presented below, using as initial state the same as that shown in \figurename{\ref{fig:scenario_initial}}. The final results after 120 ticks are summarized in \figurename{\ref{fig:auto_plots}}, showing through the plots the relationship between mission risk, compromised nodes, conflict states, and telemetry events. Initially, during ticks 1--22, events are observed that indicate the compromise of nodes in the TEZ (starting with \texttt{TEZ-GW}), whose impact on mission risk is minimal (\(R=0.5\)). However, the appearance of compromised nodes is sufficient to escalate the system to conflict state S1, and the defense policy applies \texttt{A6-Contain-TEZ} to protect key tactical-edge nodes.
At tick 22, the compromise of \texttt{TEZ-B2} causes escalation to \(S_2\) (the MOZ is now reachable), after which the policy switches to \texttt{A1-Quarantine-Node} over the TEZ frontier. After a few ticks, all nodes in the TEZ become compromised. Then, between ticks 34--55, repeated \texttt{PRE@MOZ-IN} alerts appear, but the actual pivot into MOZ is delayed until tick 56.  
From tick 56 onward, compromised nodes begin to be identified in the MOZ \texttt{MOZ-IN} and \texttt{RELAY}, which pose a higher operational risk, and the defense recommendation changes to \texttt{A5-Protect-Critical}. However, the protections are not sufficient to avoid compromise, and it is reflected in an abnormal increase in mission risk, therefore a rapid transition from state S2 to S5 in a very short period of time. The compromise of \texttt{ISR} and \texttt{C2} at ticks 64--65 escalates the state to \(S_5\) and increases the mission-risk score from \(7.34\) to \(12.39\). Later compromise of \texttt{COP}, \texttt{MOZ-B}, and \texttt{ESZ-IN} (tick 96 onward) further raises the risk to \(20.43\).  Although not fully reflected in the plots, the automatically selected defensive actions influence the evolution of the simulation by increasing defensive capabilities and reducing the probability of node compromise. In fact, the application of the A1-Quarantine-Node measures (after reaching S2 at tick 22) helps prevent a direct transition to the MOZ until 34 ticks into the execution.

Figure~\ref{fig:auto_plots} summarizes the automatic execution, showing that mission risk remains low during the early TEZ phase, increases after MOZ compromise, and rises sharply once mission-critical assets are affected, while the compromised-nodes plot reflects staged expansion across zones, the conflict-state plot captures escalation from initial compromise to mission impact, and the observed-events plot highlights the partial and time-varying nature of telemetry. Overall, the results indicate a structured, directional attacker progression from TEZ to frontier nodes, into MOZ, and eventually toward ESZ and mission-critical assets, and clear modulation of attack dynamics by defensive actions, where \texttt{A6} delays early spread, \texttt{A1} supports frontier containment, and \texttt{A5} prioritizes protection of mission-critical assets.

\subsection{Limitations}
\label{subsec:integrated_validation}

The validation presented in this work is scenario-based and illustrative. The NetLogo prototype demonstrates that the proposed framework can be instantiated as an executable cyber-defense environment. However, the current validation should not be interpreted as empirical evidence of operational effectiveness. The scenario is controlled, the topology is synthetic, and the reported execution corresponds to a representative trace rather than to a statistically exhaustive set of experiments. The prototype simplifies several elements of the mathematical model: conflict states are derived from explicit node compromise rather than posterior beliefs, and the automatic response mechanism relies on a heuristic policy instead of a globally optimal or multi-objective optimization procedure. Nevertheless, these limitations do not invalidate the objective of the prototype as a proof of concept. Together with the analytical properties discussed in Section~\ref{subsec:mathematical_consistency_robustness}, the simulation results support the internal coherence of the proposal: the framework can be instantiated in an executable environment where progression, telemetry, conflict states, mission-risk, and defensive responses remain aligned with the intended mission-oriented cyber-defense loop.

\section{Conclusions and Future Work}
\label{sec:conclusions}
This paper presented a mission-oriented cyber-defense CSA framework that connects an integrated three-layer probabilistic model with an executable NetLogo simulation prototype. A central contribution of the proposal is the explicit treatment of uncertainty from the defender's perspective. The true adversarial state is not assumed to be directly observable; instead, attacker progression is represented through hidden variables, partial and noisy telemetry is transformed into posterior compromise beliefs, and these beliefs are used to support conflict-state abstraction, mission-risk assessment, and defensive action selection. In this sense, the framework contributes to CSA by linking perception and comprehension through telemetry-driven inference, projection through mission-risk estimation, and decision support through a one-step rule that balances residual mission risk and operational cost. The validation results support the internal coherence of the proposal. The mathematical layer provides well-posed inference and local robustness under bounded perturbations of the observation model. The simulation results show staged attacker progression, coherent telemetry, meaningful conflict-state escalation, and mission-risk increases aligned with the compromise of mission-critical assets such as \texttt{ISR}, \texttt{C2}, and \texttt{COP}.

Future work should extend the validation methodology in two directions. First, scenario realism should be improved by using larger and more operationally grounded cyber-defense graphs, including topologies derived from cyber-range exercises, mission-thread analysis, or MITRE ATT\&CK/D3FEND-informed adversarial and defensive mappings. Second, the experimental methodology should become more quantitative by executing batches of stochastic simulations under fixed random seeds, reporting aggregate metrics such as time to mission impact, maximum and cumulative mission risk, number of compromised mission-critical assets, conflict-state stability, and action distribution. 

\begin{credits}
\subsubsection{\ackname} \tiny EU \normalsize This work has been co-funded by the European Union (EDF program; project ECYSAP EYE). Views and opinions expressed are however those of the author(s) only and do not necessarily reflect those of the European Union or the European Defence Fund. Neither the European Union nor the granting authority can be held responsible for them.
\subsubsection{Disclosure of Interests.} The authors have no competing interests to declare that are relevant to the content of this article.
\end{credits}

\bibliographystyle{splncs04}
\bibliography{bibliography}

@inproceedings{phillips1998graphbased,
  author    = {Phillips, Cynthia and Swiler, Laura Painton},
  title     = {A graph-based system for network-vulnerability analysis},
  booktitle = {Proceedings of the 1998 Workshop on New Security Paradigms},
  year      = {1998},
  pages     = {71--79},
  doi       = {10.1145/310889.310919}
}

@inproceedings{ou2005mulval,
  author    = {Ou, Xinming and Govindavajhala, Sudhakar and Appel, Andrew W.},
  title     = {MulVAL: A Logic-based Network Security Analyzer},
  booktitle = {Proceedings of the 14th USENIX Security Symposium},
  year      = {2005},
  pages     = {113--128}
}

@inproceedings{NetSPA,
  author    = {Ingols, Kyle and Lippmann, Richard and Piwowarski, Keith},
  title     = {Practical Attack Graph Generation for Network Defense},
  booktitle = {Proceedings of the 22nd Annual Computer Security Applications Conference (ACSAC)},
  year      = {2006},
  pages     = {121--130},
  doi       = {10.1109/ACSAC.2006.39}
}

@article{MulVALSurvey,
  author  = {Tayouri, David and Baum, Nick and Shabtai, Asaf and Puzis, Rami},
  title   = {A Survey of MulVAL Extensions and Their Attack Scenarios Coverage},
  journal = {IEEE Access},
  year    = {2023},
  volume  = {11},
  pages   = {27974--27991},
  doi     = {10.1109/ACCESS.2023.3257721}
}

@article{ADT_JLC,
  author  = {Kordy, Barbara and Mauw, Sjouke and Radomirovi{\'c}, Sa{\v{s}}a and Schweitzer, Patrick},
  title   = {Attack-defense Trees},
  journal = {Journal of Logic and Computation},
  year    = {2014},
  doi     = {10.1093/logcom/exs029}
}

@article{munoz2019,
  author  = {Mu{\~n}oz-Gonz{\'a}lez, Luis and Sgandurra, Daniele and Barr{\`e}re, Mart{\'i}n and Lupu, Emil C.},
  title   = {Exact Inference Techniques for the Analysis of Bayesian Attack Graphs},
  journal = {IEEE Transactions on Dependable and Secure Computing},
  volume  = {16},
  number  = {2},
  pages   = {231--244},
  year    = {2019},
  doi     = {10.1109/TDSC.2016.2627033}
}

@mastersthesis{frigault2010,
  author = {Frigault, Marcel},
  title  = {Measuring Network Security Using Bayesian Network-Based Attack Graphs},
  school = {Concordia University, Concordia Institute for Information Systems Engineering},
  year   = {2010}
}

@inproceedings{frigault2008,
  author    = {Frigault, Marcel and Wang, Lingyu and Singhal, Anoop and Jajodia, Sushil},
  title     = {Measuring Network Security Using Dynamic Bayesian Network},
  booktitle = {Proceedings of the 4th ACM Workshop on Quality of Protection (QoP '08)},
  pages     = {23--30},
  year      = {2008},
  publisher = {ACM},
  doi       = {10.1145/1456362.1456368}
}

@article{BAG,
  author  = {Poolsappasit, Nayot and Dewri, Rinku and Ray, Indrajit},
  title   = {Dynamic Security Risk Management Using Bayesian Attack Graphs},
  journal = {IEEE Transactions on Dependable and Secure Computing},
  volume  = {9},
  number  = {1},
  pages   = {61--74},
  year    = {2012},
  doi     = {10.1109/TDSC.2011.34}
}

@article{BAG2,
  author  = {Javorn{\'i}k, Martin and Hus{\'a}k, Martin},
  title   = {Mission-centric decision support in cybersecurity via Bayesian Privilege Attack Graph},
  journal = {Engineering Reports},
  volume  = {4},
  pages   = {e12538},
  year    = {2022},
  doi     = {10.1002/eng2.12538}
}

@inproceedings{BAG3,
  author    = {Ahmadian Ramaki, Ali and Khosravi-Farmad, Mohammad and Ghaemi Bafghi, Amin},
  title     = {Real Time Alert Correlation and Prediction using Bayesian Networks},
  booktitle = {2015 12th International Iranian Society of Cryptology Conference on Information Security and Cryptology (ISCISC)},
  year      = {2015},
  doi       = {10.1109/ISCISC.2015.7387905}
}

@article{HMM1,
  author  = {Ghafir, Ibrahim and Kyriakopoulos, Konstantinos G. and Lambotharan, Sangarapillai and Aparicio-Navarro, Francisco J. and Assadhan, Basil and BinSalleeh, Hamad and Diab, Diab M.},
  title   = {HMMs and Alert Correlations for the Prediction of APTs},
  journal = {IEEE Access},
  volume  = {7},
  pages   = {99508--99520},
  year    = {2019},
  doi     = {10.1109/ACCESS.2019.2930200}
}

@article{zhang2022msahmm,
  author  = {Zhang, Xu and Wu, Ting and Zheng, Qiuhua and Zhai, Liang and Hu, Haizhong and Yin, Weihao and Zeng, Yingpei and Cheng, Chuanhui},
  title   = {Multi-Step Attack Detection Based on Pre-Trained Hidden Markov Models},
  journal = {Sensors},
  year    = {2022},
  volume  = {22},
  number  = {8},
  pages   = {2874},
  doi     = {10.3390/s22082874}
}

@article{POMDP2,
  author  = {Miehling, Erik and Rasouli, Mohamad and Teneketzis, Demosthenis},
  title   = {A POMDP Approach to the Dynamic Defense of Large-Scale Cyber Networks},
  journal = {IEEE Transactions on Information Forensics and Security},
  volume  = {13},
  number  = {10},
  pages   = {2490--2505},
  year    = {2018},
  doi     = {10.1109/TIFS.2018.2819967}
}

@article{POMDP1,
  author  = {Hu, Zhisheng and Zhu, Minghui and Liu, Peng},
  title   = {Adaptive Cyber Defense Against Multi-Stage Attacks Using Learning-Based POMDP},
  journal = {ACM Transactions on Privacy and Security},
  volume  = {24},
  number  = {1},
  year    = {2020},
  doi     = {10.1145/3418897}
}

@article{POMDP3,
  author  = {Kazeminajafabadi, Armita and Imani, Mahdi},
  title   = {Optimal Joint Defense and Monitoring for Networks Security under Uncertainty: A POMDP-Based Approach},
  journal = {IET Information Security},
  year    = {2024},
  doi     = {10.1049/2024/7966713}
}

@article{monahan1982state,
  author  = {Monahan, George E.},
  title   = {A survey of partially observable Markov decision processes: Theory, models, and algorithms},
  journal = {Management Science},
  year    = {1982},
  volume  = {28},
  number  = {1},
  pages   = {1--16},
  doi     = {10.1287/mnsc.28.1.1}
}

@article{kaelbling1998planning,
  author  = {Kaelbling, Leslie Pack and Littman, Michael L. and Cassandra, Anthony R.},
  title   = {Planning and acting in partially observable stochastic domains},
  journal = {Artificial Intelligence},
  year    = {1998},
  volume  = {101},
  number  = {1--2},
  pages   = {99--134},
  doi     = {10.1016/S0004-3702(98)00023-X}
}

@article{FCM,
  author  = {Poleto, T. and Carvalho, V. D. H. de and Silva, A. L. B. da and Clemente, T. R. N. and Silva, M. M. and Gusm{\~a}o, A. P. H. de and Costa, A. P. C. S. and Nepomuceno, T. C. C.},
  title   = {Cybersecurity for Telehealth Scenarios: Analysis Based on a Fuzzy Cognitive Map},
  journal = {Healthcare},
  volume  = {9},
  pages   = {1504},
  year    = {2021},
  doi     = {10.3390/healthcare9111504}
}

@article{felix2019fcm_review,
  author  = {Felix, Gerardo and N{\'a}poles, Gonzalo and Falcon, Rafael and Froelich, Wojciech and Vanhoof, Koen and Bello, Rafael},
  title   = {A review on methods and software for fuzzy cognitive maps},
  journal = {Artificial Intelligence Review},
  year    = {2019},
  volume  = {52},
  pages   = {1707--1737},
  doi     = {10.1007/s10462-017-9575-1}
}

@article{GTSurveyAPT,
  author  = {Khalid, Mohd Nor Akmal and Al-Kadhimi, Amjed Ahmed and Singh, Manmeet Mahinderjit},
  title   = {Recent Developments in Game-Theory Approaches for the Detection and Defense against Advanced Persistent Threats: A Systematic Review},
  journal = {Mathematics},
  year    = {2023},
  volume  = {11},
  pages   = {1353},
  doi     = {10.3390/math11061353}
}

@article{EndsleySA,
author = {Endsley, Mica},
year = {2000},
month = {01},
pages = {3-32},
title = {Theoretical underpinnings of situation awareness: A critical review},
journal = {Situation awareness analysis and measurement}
}

@misc{NetLogo,
  author       = {Wilensky, Uri},
  title        = {{NetLogo}},
  year         = {1999},
  howpublished = {Center for Connected Learning and Computer-Based Modeling, Northwestern University, Evanston, IL},
  url          = {http://ccl.northwestern.edu/netlogo/},
  note         = {Accessed: 2026-05-14}
}

@inproceedings{javornik2019mission,
author    = {Javorn{\'i}k, Michal and Kom{\'a}rkov\'a, Jana and Hus{\'a}k, Martin},
title     = {Decision Support for Mission-Centric Cyber Defence},
booktitle = {Proceedings of the 14th International Conference on Availability, Reliability and Security (ARES 2019)},
year      = {2019},
pages     = {1--8},
publisher = {ACM},
doi       = {10.1145/3339252.3340522}
}

@article{motzek2017pareto,
author  = {Motzek, Alexander and Gonzalez-Granadillo, Gustavo and Debar, Herv{\'e} and Garcia-Alfaro, Joaquin and M{\"o}ller, Ralf},
title   = {Selection of Pareto-efficient response plans based on financial and operational assessments},
journal = {EURASIP Journal on Information Security},
year    = {2017},
volume  = {2017},
number  = {1},
pages   = {1--22},
doi     = {10.1186/s13635-017-0063-6}
}

\end{document}